\documentclass[twocolumn,floatfix,aps,prb,nofootinbib,eqsecnum]{revtex4-2}
\usepackage{graphicx}
\usepackage{amsmath}
\usepackage{amssymb}
\usepackage{bm}
\usepackage{hyperref}

\usepackage{color}

\begin{document}

\title{Chiral charge transfer along magnetic field lines in a Weyl superconductor}
\author{G. Lemut}
\affiliation{Instituut-Lorentz, Universiteit Leiden, P.O. Box 9506, 2300 RA Leiden, The Netherlands}
\author{M. J. Pacholski}
\affiliation{Instituut-Lorentz, Universiteit Leiden, P.O. Box 9506, 2300 RA Leiden, The Netherlands}
\author{C. W. J. Beenakker}
\affiliation{Instituut-Lorentz, Universiteit Leiden, P.O. Box 9506, 2300 RA Leiden, The Netherlands}
\date{June 2021}
\begin{abstract}
We identify a signature of chirality in the electrical conduction along magnetic vortices in a Weyl superconductor: The conductance depends on whether the magnetic field is parallel or antiparallel to the vector in the Brillouin zone that separates Weyl points of opposite chirality.   
\end{abstract}
\maketitle

\section{Introduction}

Three-dimensional Weyl fermions have a definite chirality, given by the $\pm$ sign in the Weyl Hamiltonian $\pm \bm{p}\cdot\bm{\sigma}$. Three spatial dimensions are essential, if $\bm{p}\cdot\bm{\sigma}=p_x\sigma_x+p_y\sigma_y$ contains only two Pauli matrices, then $+\bm{p}\cdot\bm{\sigma}$ and $-\bm{p}\cdot\bm{\sigma}$ can be transformed into each other by a unitary transformation (conjugation with $\sigma_z$). The chirality is therefore a characteristic feature of 3D Weyl semimetals, not shared by 2D graphene.

\begin{figure}[tb]
\centerline{\includegraphics[width=1\linewidth]{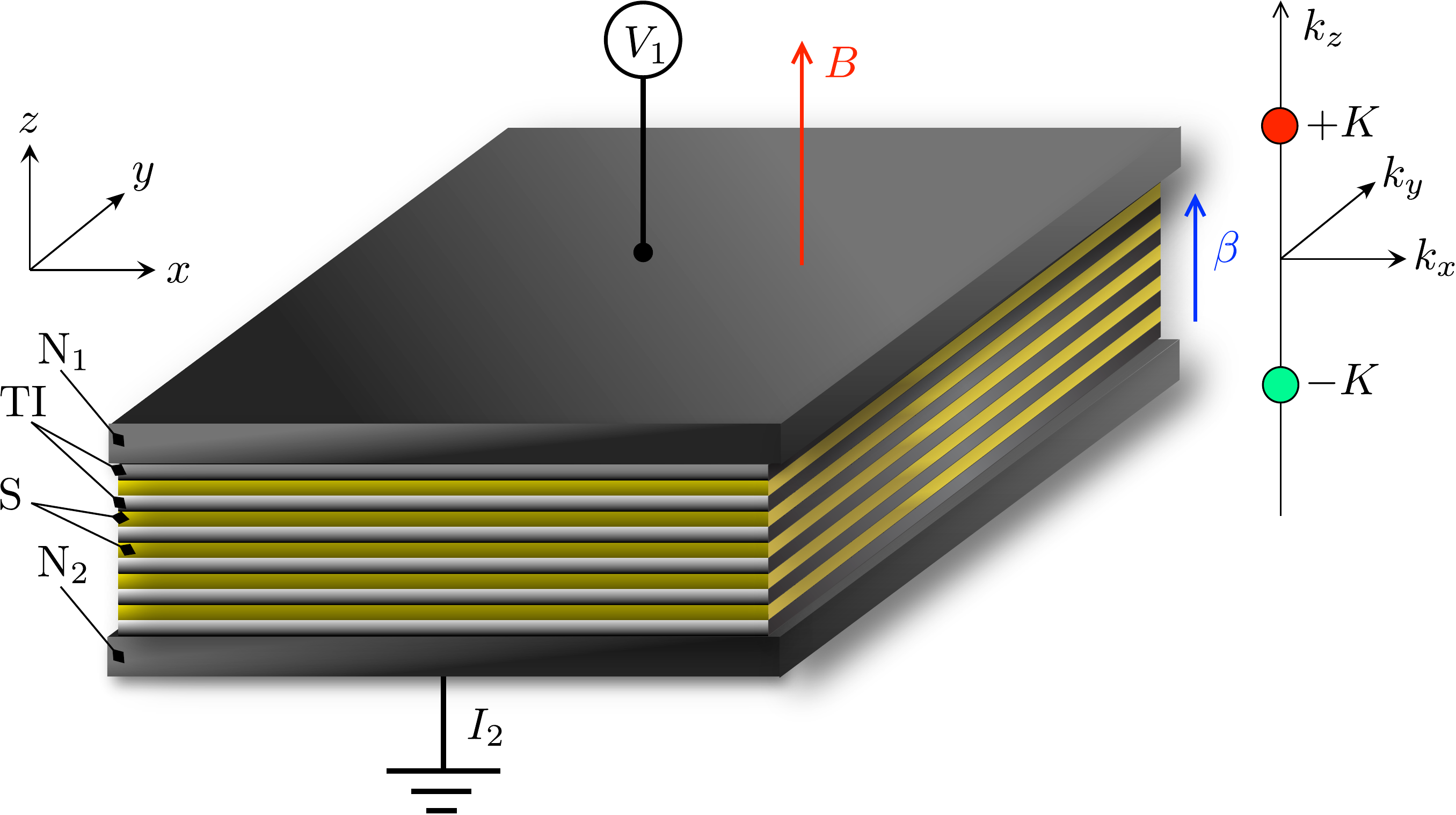}}
\caption{Left panel: Weyl superconductor formed by alternating layers of magnetic topological insulator (TI, magnetization $\beta$) and \textit{s}-wave superconductor (S, pair potential $\Delta_0$, chemical potential $\mu$), between normal-metal contacts (${\rm N}_1$ and ${\rm N}_2$. A magnetic field $B$ perpendicular to the layers (along $z$) produces a Landau band that is dispersionless in the $x$--$y$ plane, with free propagation in the $z$-direction. Right panel: Weyl points of opposite chirality at $k_z=\pm K$. For $\mu\neq 0$ the conductance $G=I_2/V_1$ depends on whether the magnetic field points parallel or antiparallel to the vector from $-K$ to $+K$. 
}
\label{fig_layout}
\end{figure}

The search for observable signatures of chirality is a common theme in the study of this new class of materials \cite{Yan17,Arm18,Ong21,Has21}. The basic mechanism used for that purpose is the chirality dependent motion in a magnetic field: Weyl fermions in the zeroth Landau level propagate parallel or antiparallel to the field lines, dependent on their chirality \cite{Nie83}. A population imbalance between the two chiralities then produces the chiral magnetic effect \cite{Kha14,Bur15}: An electrical current along the field lines, which changes sign if the field is inverted.

Here we present a novel, albeit less dramatic, signature of chirality: An electrical conductance which depends on the magnetic field direction. The effect appears if superconductivity is induced in a magnetic topological insulator, in the layered geometry of Meng and Balents \cite{Men12} (see Fig.\ \ref{fig_layout}). The superconductor cannot gap out the Weyl points of opposite chirality, provided that the induced pair potential $\Delta_0$ remains smaller than the magnetization energy $\beta$. The main effect of the superconductor is to renormalize the charge of the quasiparticles \cite{OBr17}, by a factor $\kappa=\sqrt{1-\Delta_0^2/\beta^2}$.

A magnetic field $B$ perpendicular to the layers penetrates in an array of $h/2e$ vortices. The zeroth Landau level is a dispersionless flat band in the plane of the layers --- the chirality of the Weyl fermions prevents broadening of the Landau band by vortex scattering \cite{Pac18}.

Following Ref.\ \onlinecite{Lem19} we probe the Landau band by electrical conduction: A voltage $V_1$ applied to contact ${\rm N}_1$ induces a current $I_2=GV_1$ in contact ${\rm N}_2$. This is a three-terminal circuit, the grounded superconductor being the third terminal. The chemical potential $\mu_{\rm N}$ in the normal-metal contacts is assumed to be large compared to the value $\mu$ in the superconductor. We calculate the dependence of the conductance $G(\pm B)$ on the direction of the magnetic field $B$, relative to the separation of the Weyl points of opposite chirality.

When the chemical potential is at the Weyl point ($\mu=0$) the conductance is determined by the renormalized charge and $B$ only enters via the Landau band degeneracy \cite{Lem19}, 
\begin{equation}
G=\kappa^2 G_0,\;\;G_0=(e^2/h)N_\Phi,\;\;\text{at}\;\;\mu=0,
\end{equation}
with $N_{\Phi}=eBS/h$ the flux through an area $S$ in units of $h/e$. We generalize this result to nonzero $\mu$ and find that
\begin{equation}
\delta G = G(B)-G(-B)=(4\mu/\beta)(\kappa^2-\kappa)G_0.\label{deltaGresult}
\end{equation}
The conductance thus depends on whether the magnetic field points from $+$ chirality to $-$ chirality, or the other way around.

The outline of the paper is as follows. In the next section we formulate the problem of electrical conduction along the magnetic vortices of a Weyl superconductor. The key quantity to calculate is the charge $e^\ast$ transferred by the quasiparticles across the normal-superconductor interface. At $\mu=0$ this is simply given by the renormalized charge $\kappa e$ of the Weyl fermions \cite{Lem19}, but that no longer holds at nonzero $\mu$. In Secs.\ \ref{sec_fractional} and \ref{sec_dispersion} we apply a mode matching technique developed in Ref.\ \onlinecite{Don21} to calculate $e^\ast$. The conductance then follows in Sec.\ \ref{sec_conductance}. These are all analytical results, we test them on a computer simulation of a tight-binding model in Sec.\ \ref{sec_numerics}. We conclude in Sec.\ \ref{sec_conclude}.

\section{Weyl superconductor in a magnetic vortex lattice}

We consider a three-dimensional Weyl superconductor \cite{Men12} (Fermi velocity $v_{\rm F}$, chemical potential $\mu$, \textit{s}-wave pair potential $\Delta_0 e^{i\phi}$), sandwiched between metal contacts ${\rm N}_1$ and ${\rm N}_2$ at $z=\pm L/2$ (see Fig.\ \ref{fig_layout}). A magnetic field $B>0$ in the $z$-direction penetrates the superconductor in the form of a vortex lattice. The superconducting phase $\phi$ winds by $2\pi$ around each vortex (at position $\bm{R}_n$),
\begin{equation}
\nabla\times\nabla\phi=2\pi\hat{z}\sum_n\delta(\bm{r}-\bm{R}_n).
\end{equation}

The Bogoliubov-De Gennes Hamiltonian is
\begin{align}
{\cal H}={}&v_{\rm F}\nu_z\tau_z(\bm{k}\cdot\bm{\sigma})-ev_{\rm F}\nu_0\tau_z(\bm{A}\cdot\bm{\sigma})+\nu_0\tau_0\bm{\beta}\cdot\bm{\sigma}\nonumber\\
&-\mu\nu_z\tau_0\sigma_0+\Delta_0(\nu_x\cos\phi-\nu_y\sin\phi)\tau_0\sigma_0.\label{calHdef}
\end{align}
The Pauli matrices $\sigma_\alpha,\tau_\alpha,\nu_\alpha$ act respectively on the spin, subband, and electron-hole degree of freedom. We set $\hbar$ to unity and choose the electron charge as $+e$. The magnetization $\bm{\beta}=\beta\bm{n}_\beta$ (with $\bm{n}_\beta$ a unit vector) may point in an arbitary direction relative to $\bm{B}=\nabla\times\bm{A}=B\hat{z}$. We choose a gauge in which $A_z=0$ and both $\bm{A}$ and $\phi$ are $z$-independent.

The Weyl points in zero magnetic field are at momentum $\bm{k}=\pm\bm{K}=\pm K\bm{n}_\beta$ with
\begin{equation}
v_{\rm F}K=\kappa\beta,\;\;\kappa=\sqrt{1-\Delta_0^2/\beta^2}.
\end{equation}
The Weyl cones remain gapless provided that $\Delta_0<\beta$. In a magnetic field the states condense into Landau bands, dispersionless in the plane perpendicular to $\bm{B}$, but freely moving along $\bm{B}$. 

A quasiparticle in a Landau band, at energy $E$, has charge expectation value $Q=-e\partial E/\partial\mu$. At the Weyl point, $\mu=0=E$, this equals \cite{OBr17}
\begin{equation}
Q_0=\kappa e=e\sqrt{1-\Delta_0^2/\beta^2}.
\end{equation}
We seek the charge $e^\ast$ transferred into the normal-metal contact by a quasiparticle in the Landau band. At $\mu=0$ this was calculated in Ref.\ \onlinecite{Lem19}, with the result $e^\ast=Q_0$. We wish to generalize this to nonzero  $\mu$. For that purpose we apply a methodology developed for a different problem in Ref.\ \onlinecite{Don21}, as described in the next section.

\section{Fractional charge transfer}
\label{sec_fractional}

\subsection{Matching condition}

The particle current operator $\hat{v}_z$ and charge current operator $\hat{j}_z$, both in the $z$-direction, are given by
\begin{equation}
\begin{split}
&\hat{v}_z=\partial{\cal H}/\partial k_z=v_{\rm F}\nu_z\tau_z\sigma_z,\\
&\hat{j}_z=-\partial{\cal H}/\partial A_z=ev_{\rm F}\nu_0\tau_z\sigma_z.
\end{split}
\end{equation}
In what follows we set $v_{\rm F}$ and $e$ equal to unity, for ease of notation.

The chirality $\chi=\pm 1$ of a mode in the superconductor (S) determines whether it propagates in the $+z$ direction or in the $-z$ direction. We position the normal-superconductor (NS) interface at $z=0$, so that the mode in $S$ approaches it from $z<0$ for $\chi=+1$ and from $z>0$ for $\chi=-1$. 

We assume that the chemical potential $\mu_{\rm N}$ in N is large compared to the value $\mu$ in S. The potential step at the NS interface boosts the momentum component $k_z$ perpendicular to the interface, without affecting the parallel components $k_x,k_y$, so in N only modes are excited with $|k_z|\gg |k_x|,|k_y|$. These are eigenstates of $\nu_z\tau_z\sigma_z$ with eigenvalue $\chi$, moving away from the interface in the $+z$ direction if $\chi=+1$ and in the $-z$ direction for $\chi=-1$. Continuity of the wave function $\Psi$ at the interface then gives the matching condition
\begin{equation}
\nu_z\tau_z\sigma_z\Psi=\chi\Psi\;\;\text{at}\;\;z=0.\label{boundarycond}
\end{equation}

\subsection{Projection}

Because the Hamiltonian \eqref{calHdef} commutes with $\tau_z$ we can replace this Pauli matrix by the subband index $\tau=\pm 1$ and rewrite the matching condition \eqref{boundarycond} as $\chi\tau\nu_z\sigma_z\Psi=\Psi$. We define the projection operator
\begin{equation}
{\cal P}=\tfrac{1}{2}(1+\chi\tau\nu_z\sigma_z),\;\;\text{such that}\;\;{\cal P}\Psi=\Psi\;\;\text{at}\;\;z=0,\label{Pmatching}
\end{equation}
and project the Hamiltonian \eqref{calHdef},
\begin{equation}
{\cal P}{\cal H}{\cal P}=(\tau\beta_z-\chi\mu){\cal P}\hat{j}_z{\cal P}+{\cal P}\hat{k}_z\hat{v}_z{\cal P}.\label{calHdefprojected}
\end{equation}
We have used that $\bm{A}$ only has components in the $x$--$y$ plane. The hat on $\hat{k}_z=-i\partial/\partial z$ is there to remind us it is an operator.
 
 We take the $z$-dependent inner product
\begin{equation}
\langle\Psi_1|\Psi_2\rangle_z=\int dx\int dy\, \Psi_2^\ast(x,y,z)\Psi_2(x,y,z)
\end{equation}
 of Eq.\ \eqref{calHdefprojected},
\begin{equation}
\begin{split}
&(\tau\beta_z-\chi\mu)\langle\Psi|{\cal P}\hat{j}_z{\cal P}|\Psi\rangle_z=\langle\Psi|{\cal P}\delta{\cal H}{\cal P}|\Psi\rangle_z,\\
&\text{with}\;\;\delta{\cal H}={\cal H}-\hat{k}_z\hat{v}_z.
\end{split}
\end{equation}

At the NS interface $z=0$ the projector may be removed,
\begin{equation}
(\tau\beta_z-\chi\mu)\langle\Psi|\hat{j}_z|\Psi\rangle_0=\langle\Psi|\delta{\cal H}|\Psi\rangle_0, \label{jzvzb}
\end{equation}
since neither $\hat{j}_z$ nor $\delta{\cal H}$ contain a $z$-derivative, so that these operators commute with the limit $z\rightarrow 0$ and we may replace ${\cal P}\Psi$ by $\Psi$ in view of the matching condition \eqref{Pmatching}. Eq.\ \eqref{jzvzb} is the key identity that allows us to calculate the transferred charge.

\subsection{Transferred charge}

Let $\Psi$ be an eigenstate of ${\cal H}$ at energy $E$. The transferred charge $e^\ast$ through the NS interface is given by the ratio
\begin{equation}
e^\ast=\frac{\langle\Psi|\hat{j}_z|\Psi\rangle_0}{\langle\Psi|\hat{v}_z|\Psi\rangle_0}.
\end{equation}
Substitution of Eq.\ \eqref{jzvzb} equates this to
\begin{subequations}
\begin{align}
e^\ast&=(\tau\beta_z-\chi\mu)^{-1}\frac{\langle\Psi|{\cal H}-\hat{k}_z\hat{v}_z|\Psi\rangle_0}{\langle\Psi|\hat{v}_z|\Psi\rangle_0}\label{east1}\\
&=(\tau\beta_z-\chi\mu)^{-1}\left(\chi E-\frac{\langle\Psi|\hat{k}_z\hat{v}_z|\Psi\rangle_0}{\langle\Psi|\hat{v}_z|\Psi\rangle_0}\right).\label{east2}
\end{align}
\end{subequations}
The term $\chi E$ appears because
\begin{equation}
\langle\Psi|{\cal H}|\Psi\rangle_0=E\langle\Psi|\Psi\rangle_0=\chi E\langle\Psi|\hat{v}_z|\Psi\rangle_0,
\end{equation}
where in the last equality we used the matching condition \eqref{boundarycond}.

Particle current conservation requires that 
\begin{equation}
\frac{d}{dz}\langle\Psi|\hat{v}_z|\Psi\rangle_z=0.\label{eq_ddz1}
\end{equation}
More generally, for our case of a $z$-independent Hamiltonian it holds that
\begin{equation}
\frac{d}{dz}\langle\Psi|f(\hat{k}_z)\hat{v}_z|\Psi\rangle_z=0\label{eq_ddz2}
\end{equation}
for any function of $f$ of $\hat{k}_z$ (see App.\ \ref{app_ddz} for a proof). Each of the two expectation values $\langle\cdots\rangle_0$ on the right-hand-side of Eq.\ \eqref{east2} can thus be replaced by $\langle\cdots\rangle_z$. This ratio can then be evaluated for large $|z|$, far from the NS interface, where evanescent waves have decayed and $\Psi\propto e^{ik_z z}$ is an eigenstate of $\hat{k}_z$. 

We finally obtain the transferred charge
\begin{equation}
e^\ast=e\frac{\chi E-v_{\rm F}k_z}{\tau\beta_z-\chi\mu},\label{eastkz}
\end{equation}
reinstating units of $e$ and $v_{\rm F}$. For $\mu=0=E$, $\beta=\beta_z$, $k_z=K=\kappa\beta/v_{\rm F}$ we recover the result $e^\ast=\pm\kappa e=\pm Q_0$ of Ref.\ \onlinecite{Lem19}.

It remains to relate the momentum $k_z$ of a propagating mode at the Fermi level to the parameters of the Weyl superconductor. For that we need the dispersion relation $E(k_z)$ of the Landau band, which we calculate in the next section.

\section{Dispersion relation of the Landau band}
\label{sec_dispersion}

\subsection{Block diagonalization}

We calculate the dispersion relation of the Landau band by means of the block diagonalization approach of Ref.\ \onlinecite{Pac18}. Starting from the BdG Hamiltonian \eqref{calHdef} we first make the Anderson gauge transformation  \cite{And98}
\begin{equation}
 {\cal H}\mapsto \Omega^\dagger {\cal H}\Omega,\;\;\text{with}\;\;\Omega=\begin{pmatrix}
e^{i\phi}&0\\
0&1
\end{pmatrix}.
\end{equation}
The subblocks of $\Omega$ refer to the electron-hole $(\nu_\alpha)$ degree of freedom. The resulting Hamiltonian is
\begin{align}
{\cal H}={}&\nu_z\tau_z(\bm{k}+\bm{a})\cdot\bm{\sigma}+\nu_0\tau_z\bm{q}\cdot\bm{\sigma}+\nu_0\tau_0\bm{\beta}\cdot\bm{\sigma}\nonumber\\
&-\mu\nu_z\tau_0\sigma_0+\Delta_0\nu_x\tau_0\sigma_0,\label{calHdef2}\\
\bm{a}={}&\tfrac{1}{2}\nabla\phi,\;\;\bm{q}=\tfrac{1}{2}\nabla\phi-\bm{A}.
\end{align}
Both fields $\bm{a}$ and $\bm{q}$ have only components in the $x$--$y$ plane and are $z$-independent.

To focus on states near $\bm{K}$ we set $\bm{k}=\kappa\bm{\beta}+\delta \bm{k}$ and consider $\delta \bm{k}$ small. The component parallel to $\bm{\beta}$ of a vector $\bm{v}$ is denoted by $v_\parallel=\bm{v}\cdot\bm{n}_\beta$.

One more unitary transformation ${\cal H}\mapsto U^\dagger {\cal H}U$ with
\begin{equation}
\begin{split}
&U=\sigma_\parallel\exp\bigl(\tfrac{1}{2}i\alpha\nu_y\tau_z\sigma_\parallel\bigr),\\
&\tan\alpha=-\frac{\Delta_0}{K},\;\;\cos\alpha=-(1+\Delta_0^2/K^2)^{-1/2}=-\kappa,
\end{split}\label{Vdef} 
\end{equation}
followed by a projection onto the $\nu=\tau=\pm 1$ blocks, gives a pair of $2\times 2$ low-energy Hamiltonians,
\begin{align}
H_{\tau}={}&\tau\kappa\mu\sigma_0-(\delta\bm{k}+\bm{a}-\tau\kappa\bm{q})\cdot\bm{\sigma}\nonumber\\
& +(1-\kappa)(\delta k_\parallel+ a_\parallel +\tau q_\parallel)\sigma_\parallel.\label{Hblockdiagonal}
\end{align}

Eq.\ \eqref{Hblockdiagonal} is an anisotropic Dirac Hamiltonian, the velocity parallel to the magnetization is reduced by a factor $\kappa$. The same factor renormalizes the quasiparticle charge,
\begin{equation}
Q=-e\frac{\partial H_\tau}{\partial\mu}=-e\tau\kappa.
\end{equation}
The two Hamiltonians $H_\tau=H_\pm$ near $\bm{k}=\bm{K}$ thus describe quasiparticles of opposite charge. Another pair of oppositely charged Weyl cones exists near $\bm{k}=-\bm{K}$.

If $\bm{\beta}=(\beta\sin\theta,0, \beta\cos\theta)$ makes an angle $\theta$ with the magnetic field we have
\begin{widetext}
\begin{align}
H_{\tau}={}&\tau\kappa\mu\sigma_0-\sum_{\alpha=x,y}(\delta k_\alpha+a_\alpha-\tau \kappa q_\alpha)\sigma_\alpha-\delta k_z\sigma_z\nonumber\\
&+(1-\kappa)(\delta k_x\sin\theta+\delta k_z\cos\theta+ a_x\sin\theta+\tau q_x\sin\theta)(\sigma_x\sin\theta+\sigma_z\cos\theta),\label{Hblockdiagonal2}
\end{align}
where we used that $a_z=0=q_z$.
\end{widetext}

\subsection{Zeroth Landau band}

A major simplification appears if the magnetization $\bm{\beta}$ and the magnetic field $\bm{B}$ are either parallel or perpendicular, so $\cos\theta\equiv\gamma\in\{0,\pm 1\}$. In these cases the Hamiltonian \eqref{Hblockdiagonal2} anticommutes with $\sigma_z$ when $\mu=0=\delta k_z$. This socalled chiral symmetry implies that the zeroth Landau band is an eigenstate of $\sigma_z$, with eigenvalue $-\tau$ \cite{Pac18}. The dispersion relation then follows immediately,
\begin{align}
&E(k_z)=\tau\kappa\mu+\tau\delta k_z[1-(1-\kappa)\gamma^2]\nonumber\\
&\quad=\chi\kappa\mu+\chi(k_z-\kappa\beta\gamma)[1-(1-\kappa)\gamma^2].
\end{align}
In the second equation we have identified the chirality index $\chi\equiv\text{sign}\,(dE/dk_z)=\tau$.

Equating $E(k_z)=E$ and solving for $k_z$ gives
\begin{equation}
k_z=\kappa\beta\gamma-\frac{\kappa\mu-\chi E}{1-(1-\kappa)\gamma^2},\label{kzresult}
\end{equation}
to first order in $E$ and $\mu$. (Higher order terms are not captured by the linearization around the Weyl point.)

We substitute Eq.\ \eqref{kzresult} in the expression \eqref{eastkz} for the transferred charge,
\begin{equation}
e^\ast=\frac{\chi e}{\mu-\beta\gamma}\left(\kappa\beta\gamma-\frac{\kappa\mu-\chi E}{1-(1-\kappa)\gamma^2}-\chi E\right).\label{eastcostheta}
\end{equation}
For $\bm{\beta}\parallel\bm{B}$ this gives
\begin{equation}
e^\ast=-\chi e\,\frac{\pm\kappa\beta-\mu+\chi E(1/\kappa-1)}{\pm\beta-\mu},\;\;\gamma=\pm 1.
\end{equation}
In contrast, for $\bm{\beta}\perp\bm{B}$ the $\mu$ and $E$ dependence drops out,
\begin{equation}
e^\ast=-\chi\kappa e,\;\;\gamma=0.
\end{equation}

These are the results for the charge transferred by a mode with $k_z$ near $+K$. The mode with $k_z$ near $-K$ is its charge-conjugate, the transferred charge is given by $e^\ast(E)\mapsto -e^{\ast}(-E)$.

\subsection{Comparison of transferred charge and charge expectation value}

For the case $\chi=1$ that $\bm{\beta}$ is parallel to $\bm{B}$ we can use the more accurate dispersion relation from Ref.\ \onlinecite{Pac18}, without making the linearization around the Weyl point:
\begin{equation}
E(k_z)=-\chi M(k_z)-\chi M'(k_z)\mu,\;\;M(k_z)=\beta-\sqrt{\Delta_0^2+k_z^2}.\label{Ekzdispersion}
\end{equation}
The solution $k_z=k_0(\mu)$ of the equation $E(k_z)=0$ then gives the transferred charge at the Fermi level ($E=0$) via
\begin{equation}
e^\ast=-\chi e\frac{k_0(\mu)}{\beta-\mu}.\label{eastfull}
\end{equation}
As a check for the linearization, to first order we find
\begin{equation}
k_0(\mu)=\sqrt{\beta^2-\Delta_0^2}-\mu+{\cal O}(\mu^2),
\end{equation}
in agreement with Eq.\ \eqref{kzresult} for $E=0$, $\gamma=1$. We checked that higher order terms are relatively insignificant for $|\mu/\beta|\lesssim 0.1$.

\begin{figure}[tb]
\centerline{\includegraphics[width=0.8\linewidth]{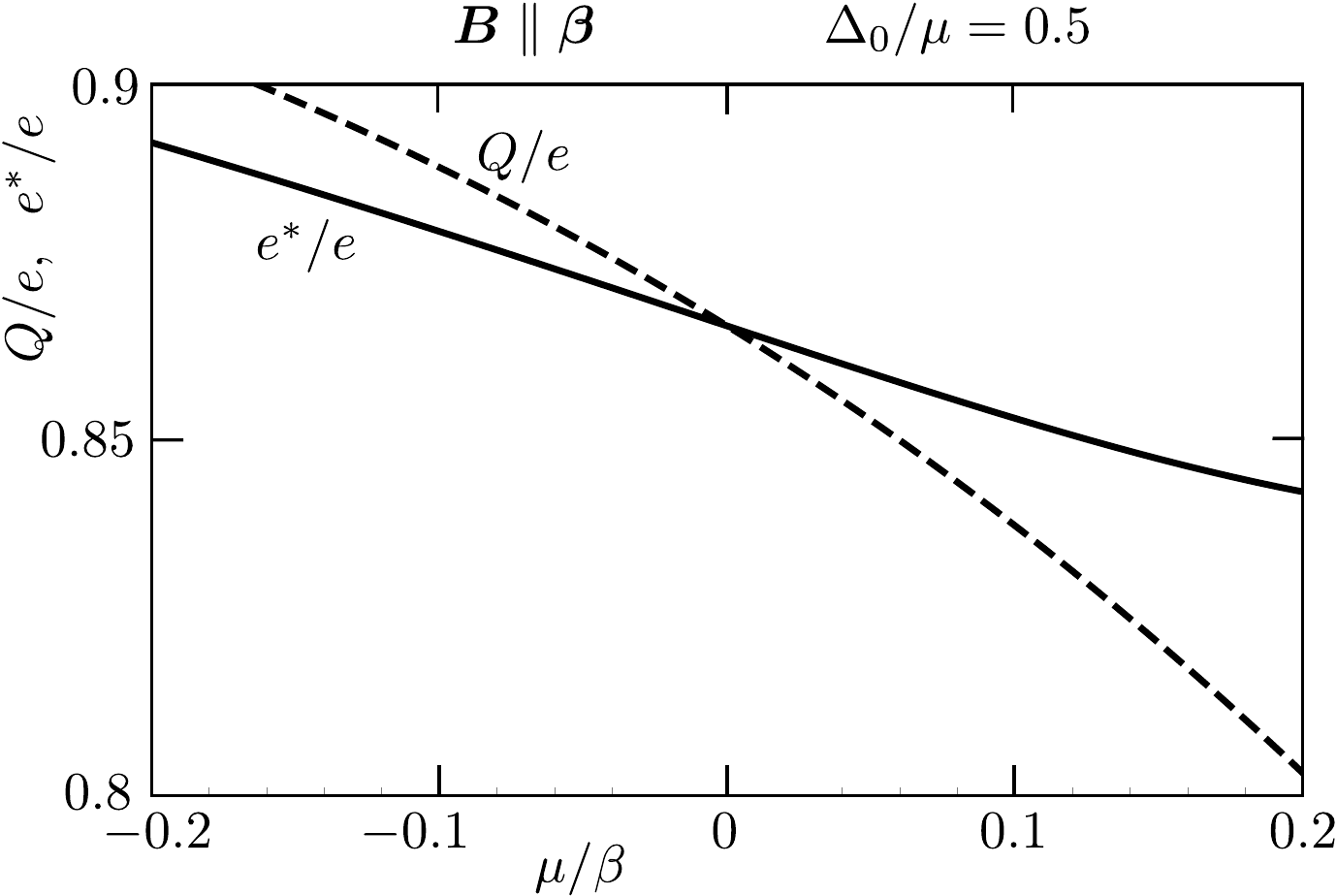}}
\caption{Comparison of the transferred charge $e^\ast$ across the NS interface and the charge expectation value $Q$ of the Weyl fermions. The curves are computed from Eqs.\ \eqref{eastfull} and \eqref{Qlinearizeda} using $k_0$ from the full nonlinear dispersion \eqref{Ekzdispersion}.
}
\label{fig_comparison}
\end{figure}

The resulting transferred charge
\begin{equation}
e^\ast=-\chi e\kappa\bigl[1+(\mu/\beta)(1-1/\kappa)+{\cal O}(\mu^2)\bigr]\label{eastlinearized}
\end{equation}
can be compared with the charge expectation value
\begin{subequations}
\begin{align}
Q&=\chi eM'(k_0)=-\frac{\chi e k_0}{\sqrt{\Delta_0^2+k_0^2}}\label{Qlinearizeda}\\
&=-\chi e\kappa\bigl[1+(\mu/\beta)(\kappa-1/\kappa)+{\cal O}(\mu^2)\bigr],\label{Qlinearizedb}
\end{align}
\end{subequations}
see Fig.\ \ref{fig_comparison}. We conclude that the $\mu$-dependence of the transferred charge $e^\ast$ is not simply accounted for by the $\mu$-dependence of the charge expectation value $Q$.

\section{Conductance}
\label{sec_conductance}

\subsection{Transmission matrix}

The Landau band contains $N_\Phi= eBS/h$ modes propagating along the magnetic field through a cross-sectional area $S$.  For each of these modes the transmission matrix $t(E)$ at energy $E$ from contact ${\rm N}_1$ to ${\rm N}_2$ is a rank-two matrix of the form
\begin{equation}
t(E)=e^{ik_z L}|\Psi^+_2\rangle\langle\Psi^+_1|+e^{-ik_z L}|\Psi^-_2\rangle\langle\Psi^-_1|.\label{tsecondrank}
\end{equation}
The incoming mode $|\Psi^\pm_1\rangle$ from contact ${\rm N}_1$ is matched in S to a Landau band mode at $\pm k_z$. This chiral mode propagates over a distance $L$ to contact ${\rm N}_2$, picking up a phase $e^{\pm i k_z L}$, and is then matched to an outgoing mode $|\Psi^\pm_2\rangle$. The  matching condition gives a charge $\pm e^\ast(\pm E)$ to $\Psi^\pm_n$,
\begin{equation}
\langle\Psi^\pm_n|\nu_z|\Psi^\pm_n\rangle=\pm e^\ast(\pm E).\label{Psipmeast}
\end{equation}

The transmission matrix $t(E)$ has electron and hole submatrices $t_{ee}$ and $t_{he}$ (transmission of an electron as an electron or as a hole). These determine the differential conductance
\begin{align}
\frac{dI_2}{dV_1}&=G_0\,\lim_{E\rightarrow eV_1}\,{\rm Tr}\,\bigl(t_{ee}^\dagger t_{ee}^{\vphantom{\dagger}}-t_{he}^\dagger t_{he}^{\vphantom{\dagger}}\bigr)\nonumber\\
&=\tfrac{1}{2}G_0\,{\rm Tr}\,(1+\nu_z)t^\dagger(eV_1)\nu_z t(eV_1),\label{dI2dV1}
\end{align}
with $G_0=N_\Phi e^2/h$.

\subsection{Linear response}

The linear response conductance $G=\lim_{V_1\rightarrow 0}dI_2/dV_1$ simplifies because at the Fermi level we can use the particle-hole symmetry relations
\begin{equation}
\left.\begin{aligned}
&\nu_y\sigma_y t\nu_y\sigma_y=t^\ast\\
&|\Psi^+_n\rangle=\nu_y\sigma_y|\Psi^-_n\rangle^\ast
\end{aligned}\right\} \;\;\text{at}\;\;E=0.
\end{equation}
These two relations imply that 
\begin{equation}
\left.\begin{aligned}
&{\rm Tr}\,t^\dagger\nu_z t=0\\
&\langle\Psi^+_n|\nu_z|\Psi^-_n\rangle=0
\end{aligned}\right\} \;\;\text{at}\;\;E=0.\label{Eis0identity}
\end{equation}

The equation \eqref{dI2dV1} for the differential conductance thus reduces in linear response to
\begin{align}
G={}&\tfrac{1}{2}G_0\,{\rm Tr}\,\nu_z t^\dagger \nu_z t\nonumber\\
={}&\tfrac{1}{2}G_0\sum_{s=\pm}\,\langle\Psi^s_2|\nu_z|\Psi^s_2\rangle\langle\Psi^s_1|\nu_z|\Psi^s_1\rangle=N_\Phi\frac{(e^\ast)^2}{h}.
\end{align}
The charge $e\mapsto e^\ast$ \textit{quadratically} renormalizes the conductance \cite{Lem19}.

Application of Eq.\ \eqref{eastcostheta} at $E=0$ then gives the result
\begin{equation}
G/G_0=\begin{cases}
\kappa^2\pm(2\mu/\beta)(\kappa^2-\kappa)&\text{if}\;\;\bm{\beta}\parallel\bm{B},\\
\kappa^2&\text{if}\;\;\bm{\beta}\perp\bm{B},
\end{cases}\label{Glinearresponse}
\end{equation}
to first order in $\mu$. The $\pm$ sign refers to $\bm{\beta}$ parallel ($+$) or antiparallel ($-$) to $\bm{B}$. The difference $\delta G=G(B)-G(-B)$ is thus given by the formula \eqref{deltaGresult} announced in the introduction.

\section{Numerical results}
\label{sec_numerics}

\begin{figure*}[tb]
\centerline{\includegraphics[width=0.8\linewidth]{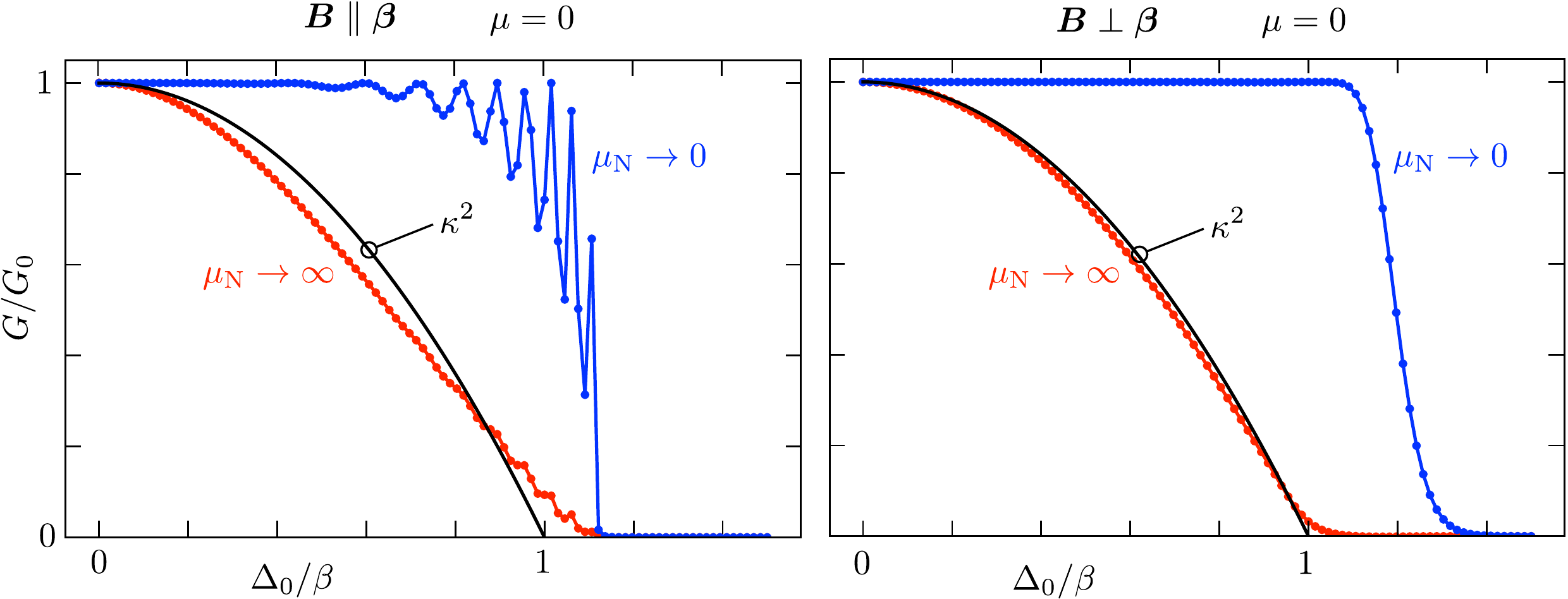}}
\caption{Dependence of the conductance $G$ on the pair potential $\Delta_0$, computed from the tight-binding model for $\bm{B}$ parallel to $\bm{\beta}$ (left panel) and for $\bm{B}$ perpendicular to $\bm{\beta}$ (right panel). The parameters are $d_0=18\,a_0$, $L=30\,a_0$, and $\mu=0$ (so there is no difference between parallel or antiparallel orientation of $\bm{B}$). The red and blue curves show the results with and without a large potential step at the NS interfaces. The black curve is the $\mu=0$ result $G=\kappa^2 G_0$ from Ref.\ \onlinecite{Lem19}.
}
\label{fig_1}
\end{figure*}

\begin{figure}[tb]
\centerline{\includegraphics[width=0.9\linewidth]{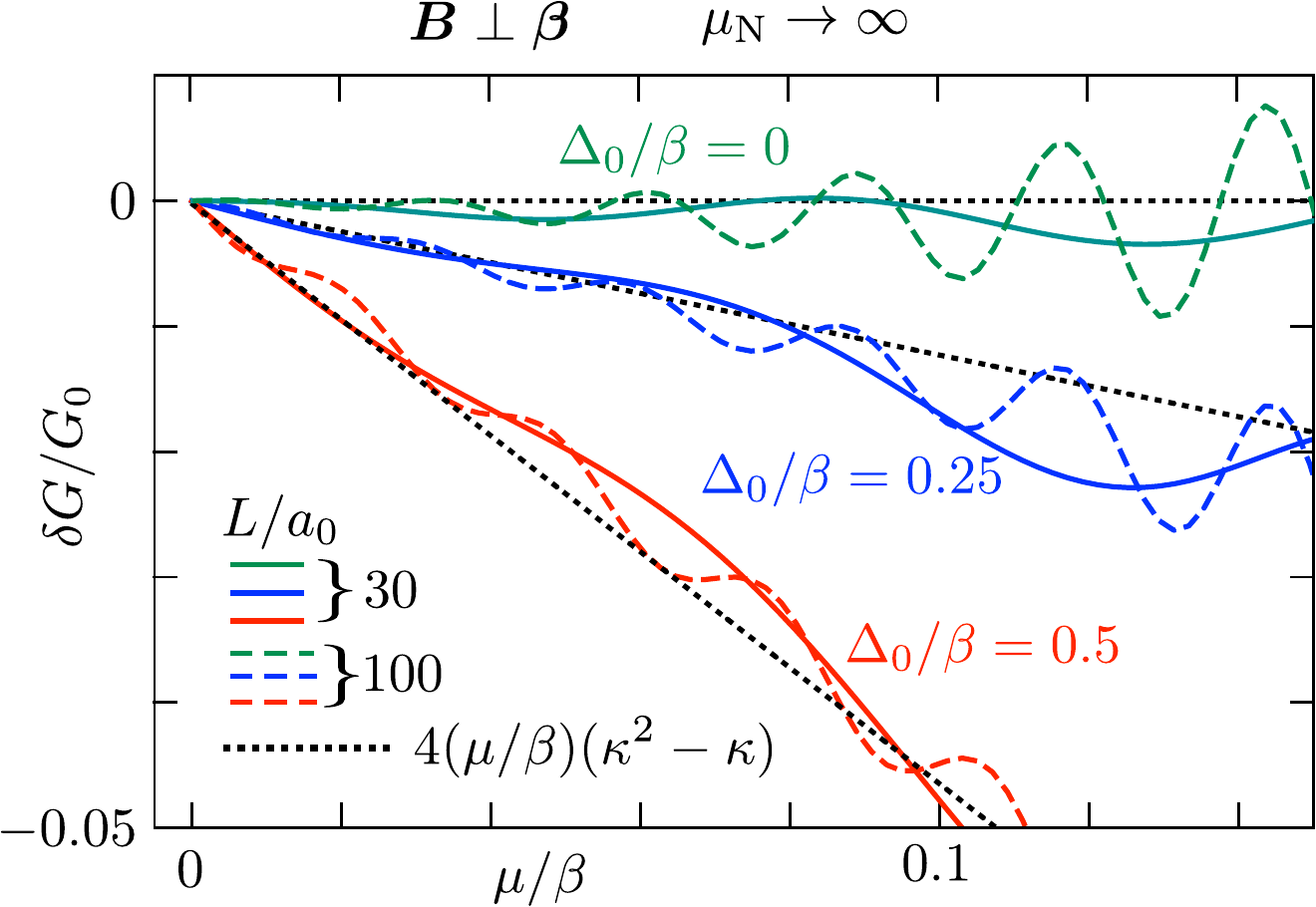}}
\caption{Dependence of the conductance on the orientation of $\bm{B}$, when it is perpendicular to $\bm{\beta}$ and there is a large potential step at the NS interfaces (limit $\mu_{\rm N}\rightarrow\infty$). The colored curves show $\delta G=G(B)-G(-B)$ as a function of $\mu$, computed from the tight-binding model ($d_0=18\,a_0$, three values of $\Delta_0/\beta$, two values of $L$). The black dotted line is the linear $\mu$-dependence following from Eq.\ \eqref{Glinearresponse}.
}
\label{fig_2}
\end{figure}

To test these analytical results, we have calculated the conductance numerically from a tight-binding model obtained by discretizing the Hamiltonian \eqref{calHdef} of the Weyl superconductor on a cubic lattice (lattice constant $a_0$):
\begin{align}
H_{\rm S}={}&(v_{\rm F}/a_0)\tau_z\sum_{\alpha=x,y,z}\sigma_\alpha\sin(a_0\nu_zk_\alpha-ea_0\nu_0A_\alpha)\nonumber\\
&+\nu_0\tau_0\bm{\beta}\cdot\bm{\sigma}-\mu\nu_z\tau_0\sigma_0\nonumber\\
&+\Delta_0(\nu_x\cos\phi-\nu_y\sin\phi)\tau_0\sigma_0\nonumber\\
&+(v_{\rm F}/a_0)\nu_z\tau_x \sigma_0\sum_{\alpha=x,y,z}(1-\cos a_0k_\alpha).\label{HdefTB}
\end{align}
The term on the last line is added to avoid fermion doubling.

The vortex lattice (a square array with lattice constant $d_0$ and two $h/2e$ vortices per unit cell) is introduced as described in Ref.\ \onlinecite{Pac18}. The scattering matrix is calculated using the Kwant code \cite{kwant}, and then the linear-response conductance follows from
\begin{equation}
G=\frac{I_2}{V_1}=\frac{e^2}{h}\,{\rm Tr}\,(t^\dagger_{ee}t_{ee}^{\vphantom{\dagger}}-t^\dagger_{he}t_{he}^{\vphantom{\dagger}}),
\end{equation}
where the trace is taken over all the $N_\Phi$ modes in the magnetic Brillouin zone and the transmission matrices are evaluated at the Fermi level ($E=0$).

In Fig.\ \ref{fig_1} we compare the conductance with and without a potential step at the NS interfaces. In the absence of a potential step, when the Hamiltonian $H_{\rm N}$ in N equals $H_{\rm S}$ with $\Delta_0=0$, the conductance has the bare value of $G_0=N_\Phi e^2/h$, as long as $\Delta_0$ remains well below $\beta$. When $\Delta_0$ exceeds $\beta$ a gap opens up at the Weyl point and the three-terminal conductance $G$ vanishes: All the carriers injected into the superconductor by contact ${\rm N}_1$ are then drained to ground before they reach contact ${\rm N}_2$.

The theory developed here does not apply to this case $\mu_{\rm N}=\mu$, but instead addresses the more realistic case $\mu_{\rm N}\gg \mu$ of a large potential step at the NS interfaces. In the numerics we implement the large-$\mu_{\rm N}$ limit by removing the transverse hoppings from the tight-binding Hamiltonian in the normal-metal leads, which is then given by
\begin{align}
H_{\rm N}={}&(v_{\rm F}/a_0)\nu_z\tau_z\sigma_z\sin a_0k_z+\nu_0\tau_0\bm{\beta}\cdot\bm{\sigma}\nonumber\\
&+(v_{\rm F}/a_0)\nu_z\tau_x \sigma_0(1-\cos a_0k_z).
\end{align}
As shown in the same Fig.\ \ref{fig_1}, in that case the conductance at $\mu=0$ follows the predicted $\kappa^2=1-\Delta_0^2/\beta^2$ parabolic profile \cite{Lem19}. The agreement is better for $\bm{B}$ perpendicular to $\bm{\beta}$ than it is for $\bm{B}$ parallel to $\bm{\beta}$.

Fig.\ \ref{fig_2} is the test of our key result, the difference \eqref{deltaGresult} of the conductance for $\bm{B}$ parallel or antiparallel to $\bm{\beta}$. The linear $\mu$-dependence has the predicted slope, without any adjustable parameter. Backscattering from the NS interfaces produces Fabry-Perot-type oscillations around this linear dependence, more rapidly oscillating when the separation $L$ of the NS interfaces is larger (compare dashed and solid curves).

\section{Conclusion}
\label{sec_conclude}

In summary, we have calculated the charge $e^\ast$ that Weyl fermions in a superconducting vortex lattice transport into a normal-metal contact. When the chemical potential $\mu$ in the superconductor is at the Weyl point, the transferred charge equals the charge expectation value $Q_0$ of the Weyl fermions \cite{Lem19} (in the limit of a large chemical potential $\mu_{\rm N}$ in the metal contacts). There is then no dependence on the relative orientation of the magnetic field $\bm{B}$ and the separation vector $\bm{\beta}$ of the Weyl points of opposite chirality. But when $\mu\neq 0$ a dependence on $\bm{B}\cdot\bm{\beta}$ appears. 

This signature of chirality shows up in the conductance, which differs if $\bm{B}$ is parallel or antiparallel to $\bm{\beta}$. It is not a large effect, a few percent (see Fig.\ \ref{fig_2}), but since it is specifically tied to the sign of the magnetic field it should stand out from other confounding effects.

We have taken a simple layered model for a Weyl superconductor \cite{Men12}, to have a definite form for the pair potential. We expect the effect to be generic for Weyl semimetals in which superconductivity is intrinsic rather than induced \cite{Bed15,Far17}. We also expect the effect to be robust to long-range disorder scattering, in view of the chirality of the motion along the magnetic field lines (backscattering needs to couple states at $\pm K$). 

\acknowledgments

This project has received funding from the Netherlands Organization for Scientific Research (NWO/OCW) and from the European Research Council (ERC) under the European Union's Horizon 2020 research and innovation programme.

\appendix

\section{Derivation of Eq. (\ref{eq_ddz2})}
\label{app_ddz}

We wish to show that the derivative
\begin{align}
&\frac{d}{dz}\langle\Psi|f(\hat{k}_z)\hat{v}_z|\Psi\rangle_z\nonumber\\
&\qquad=\langle\Psi|f(\hat{k}_z)\hat{v}_z \partial_z\Psi\rangle_z+\langle \partial_z\Psi|f(\hat{k}_z)\hat{v}_z\Psi\rangle_z\nonumber\\
&\qquad=i\langle\Psi|f(\hat{k}_z)\hat{k}_z\hat{v}_z \Psi\rangle_z-i\langle \hat{k}_z\hat{v}_z\Psi|f(\hat{k}_z)\Psi\rangle_z\label{ddzf}
\end{align}
vanishes for any function $f(\hat{k}_z)$ of $\hat{k}_z=-i\partial/\partial z$.

We rewrite
\begin{equation}
\hat{k}_z\hat{v}_z\Psi=({\cal H}-\delta{\cal H})\Psi=(E-\delta{\cal H})\Psi 
\end{equation}
and use firstly that
\begin{equation}
\langle\Psi_1|\delta{\cal H}\Psi_2\rangle_z=\langle\delta{\cal H}\Psi_1|\Psi_2\rangle_z,
\end{equation}
because $\delta{\cal H}$ does not contain any $z$-derivatives, and secondly that
\begin{equation}
[f(\hat{k}_z),\delta{\cal H}]=0,
\end{equation}
because $\delta{\cal H}$ does not depend on $z$. This gives the sequence of identities
\begin{align}
\langle\Psi|f(\hat{k}_z)\hat{k}_z\hat{v}_z\Psi\rangle_z&=\langle\Psi|f(\hat{k}_z)({\cal H}-\delta{\cal H})\Psi\rangle_z\nonumber\\
&=\langle\Psi|f(\hat{k}_z)(E-\delta{\cal H})\Psi\rangle_z\nonumber\\
&=\langle(E-\delta{\cal H})\Psi|f(\hat{k}_z)\Psi\rangle_z\nonumber\\
&=\langle({\cal H}-\delta{\cal H})\Psi|f(\hat{k}_z)\Psi\rangle_z\nonumber\\
&=\langle \hat{k}_z\hat{v}_z\Psi|f(\hat{k}_z)\Psi\rangle_z.
\end{align}

Substitution into Eq.\ \eqref{ddzf} then proves Eq.\ \eqref{eq_ddz2} from the main text.

\end{document}